# Ultrafast ultrasound coded vector Doppler imaging of blood flow velocity and resistivity[*]

YAN Shaoyuan[1], DING Yiming[1], MA Guoao[1], FU Yapeng[3], XU Kailiang[1,2], TA Dean[1,2]

1. Department of Biomedical Engineering, School of Information Science and Technology, Fudan University, Shanghai 200438, China

2. State Key Laboratory of Integrated Chips and Systems, Fudan University, Shanghai 201203, China

3. Poda Medical Technology Co., Ltd., Shanghai 200438, China

**Abstract**

Dynamic and precise measurement of cerebral blood flow velocity is crucial in neuroscience and the diagnosis of cerebrovascular diseases. Traditional color Doppler ultrasound can only measure the velocity component along the ultrasound beam, which restricts its ability to accurately capture the complete blood flow vector in complex environments. To overcome these limitations, we propose an ultrafast pulse-coded vector Doppler (PC-UVD) imaging method, utilizing Hadamard matrix-based pulse encoding to improve velocity estimation accuracy under low signal-to-noise ratio (SNR) conditions. Our study encompasses spiral flow simulations and in vivo rat brain experiments, showing significantly enhanced measurement precision compared to conventional ultrafast vector Doppler (UVD). This innovative approach enables the measurement of dynamic cerebral blood flow velocity within a single cardiac cycle, offering insights into the characteristics of cerebrovascular resistivity. The proposed PC-UVD method employs Hadamard matrix encoding of plane waves, boosting SNR without compromising temporal or spatial resolution. Velocity vectors are subsequently estimated using a weighted least squares (WLS) approach, with iterative residual-based weight optimization improving robustness to noise and minimizing the impact of outliers. The effectiveness of this technique is confirmed through simulations with a spiral blood flow phantom, demonstrating a marked improvement in velocity estimation accuracy, particularly in deep imaging regions with significant signal attenuation. In vivo experiments on rat brains further confirm that the proposed method offers greater accuracy than existing UVD approaches, particularly for small vessels. Notably, our method can precisely differentiate arterial from venous flow by analyzing pulsatility and resistivity within the cerebral vascular network. This work highlights the potential of PC-UVD for complex vascular imaging, delivering high SNR, high temporal and spatial resolution, and accurate vectorized flow measurements. Our findings underscore its capability to

---

non-invasively assess hemodynamic parameters and its potential for diagnosing cerebrovascular diseases, especially in small vessels.

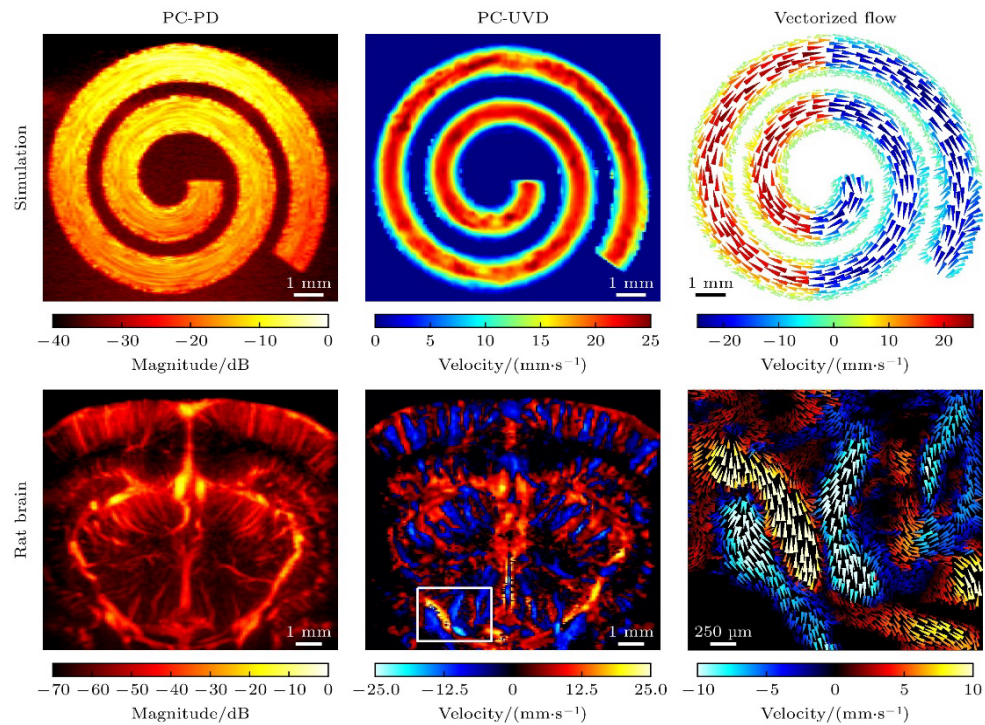

## 1. Introduction

Dynamic and accurate measurement of hemodynamic parameters, such as blood flow velocity and direction, holds significant clinical importance[1]. Furthermore, the resistivity index (RI), a key parameter for assessing the pulsatility of arteriovenous blood flow, reflects local vascular blood flow resistance and serves as a valuable indicator for indirectly evaluating vascular diseases[2]. Recent research indicates that ultrafast ultrasound Doppler imaging offers substantial potential for diagnosing cerebrovascular diseases and monitoring brain function[1,2]. Currently, the primary techniques for clinical cerebral angiography remain nuclear magnetic resonance and X-ray computed tomography, though they are limited by issues of convenience and real-time capability[3,4]. In contrast, ultrasound Doppler technology has emerged as a key tool for clinical hemodynamic assessment due to its portability, cost-effectiveness, non-invasive nature, and rapid imaging capabilities[5]; moreover, it shows significant potential for evaluating parameters such as blood flow velocity, direction, volume, and resistivity index[6].

The Doppler effect cannot detect motion perpendicular to the direction of sound wave propagation, meaning traditional color Doppler ultrasound can only measure the axial velocity component[7]. Vector Doppler imaging technology enhances traditional color Doppler imaging by acquiring projection data of true blood flow velocity from multiple angles, enabling the calculation of the blood flow velocity vector[8]. Transverse oscillation and speckle tracking are also widely used velocity vectorization imaging techniques. Transverse oscillation modulates the received signal laterally to estimate both lateral and axial motion[9], whereas speckle tracking estimates the velocity field based on the motion of the scattering intensity distribution[10]. In comparison, vector Doppler technology offers lower computational complexity and superior real-time imaging capabilities. In 1969, Fahrbach[11] calculated the vectorized velocity of a phantom at the intersection of acoustic waves using two ultrasonic probes positioned at a fixed 90° angle. In 1974, Peronneau et al.[12] extended this approach to arbitrary angles, though it remained restricted to measurements at fixed positions. In 1982, Wang and Yao[13] developed a dual-beam ultrasound vector Doppler system, which renders the beam intersection area independent of vessel position, eliminating the need for mechanical sensor adjustments based on vessel depth and expanding the measurement field of view. In 2009, Tsang et al.[14] introduced a least squares (LS)-based vector velocity estimation method, leveraging synthetic aperture technology to increase measurement angle diversity and achieving vector velocity estimation through multiple transmitting point sources and receiving apertures.

In recent years, ultrafast ultrasound imaging technology based on plane wave transmission has substantially overcome the frame rate limitations of traditional ultrasound imaging, achieving rates exceeding 10,000 frames per second[15–17]. Clutter filtering based on singular value decomposition has greatly enhanced the sensitivity of ultrafast ultrasound Doppler for detecting small blood flows[18–22]. In 2016, Yiu and Yu[23] achieved accurate measurements of multi-angle Doppler velocity projections across a wide field of view by transmitting angled plane waves, subsequently estimating the velocity vector of an in vitro blood flow phantom using the least squares method. In 2022, Wang et al.[24] coherently combined multiple angled plane waves into two intersecting beams, significantly improving velocity estimation accuracy, as validated in a carotid artery phantom experiment. In 2023, Yan et al.[25] introduced ultrafast vector Doppler (UVD) technology, integrating ultrafast ultrasound imaging with multi-angle Doppler estimation to markedly enhance blood flow imaging sensitivity, enabling high spatial and temporal resolution vectorization of small vessels and pioneering the measurement of blood flow vector velocity in the vascular networks of small animals and human brains. Despite significant advancements in vector Doppler technology in recent years, dynamic vector flow measurement in small vessels continues to face technical challenges under low SNR conditions. Vector Doppler requires retaining echo data from each angle to compute the velocity vector, preventing multi-angle coherent recombination and rendering its velocity measurements susceptible to noise, particularly in deep brain regions with low SNR and contrast, where outlier estimates are common [25]. In the linear least squares method, residuals from outliers substantially contribute to the total error, thereby significantly impairing the accuracy of velocity vector estimation for small vessels.

To address these challenges, this paper proposes a method for dynamically measuring cerebral blood flow velocity using PC-UVD. This method employs the Hadamard matrix to encode pulse polarity and decode echoes for multiple consecutive plane waves within each transmission event, thereby significantly improving the accuracy of velocity projection estimation across all angles. Subsequently, the WLS method is applied to determine the velocity vector. The WLS method adaptively mitigates the impact of noise on velocity estimation by assigning weights to each projection angle, with dynamic adjustments based on velocity estimation errors through iterative residual optimization. Specifically, the weight of projection angles with large residuals is progressively reduced, markedly enhancing the accuracy of velocity vector estimation under low signal-to-noise ratio conditions. In this study, we designed a flow simulation model featuring multiple spiral rings to analyze the feasibility of the proposed PC-UVD imaging method and quantify its measurement error through simulation experiments. Additionally, in vivo imaging of rat cerebral blood flow was conducted, comparing the accuracy of velocity vector estimation between the proposed PC-UVD method and the existing UVD method at various brain depths. Furthermore, by analyzing the variation characteristics of cerebral blood flow velocity over the cardiac cycle, the resistivity index of the entire cerebral vascular network can be estimated, enabling differentiation between arterial and venous small vessels.

## 2. Material and Methods

### 2.1 Multi-pulse plane wave imaging method based on Hadamard coding

In plane-wave transmission mode, a single plane wave pulse is emitted, and its corresponding radiofrequency (RF) echo signal is received with each transmission.

However, the velocity projection obtained from a single plane wave transmission is susceptible to noise interference, particularly in deep imaging regions where signal attenuation is pronounced. To address this, we introduce Hadamard pulse coding to improve signal quality[26].

A Hadamard matrix is a unique orthogonal matrix where each row consists of elements +1 or -1, and the dot product between any two rows is zero. For n emission angles, an n×n Hadamard matrix is employed to encode the polarity of each plane wave. The polarity of the pulse sequence transmitted in each event is dictated by a single row of the matrix. As illustrated in Fig. 1(a), +1 indicates unchanged pulse polarity, while -1 signifies a polarity reversal. For example, with n=4, the Hadamard matrix is

$$H_4 = \begin{pmatrix} +1 & +1 & +1 & +1 \\ +1 & -1 & +1 & -1 \\ +1 & +1 & -1 & -1 \\ +1 & -1 & -1 & +1 \end{pmatrix}. \quad (1)$$

As shown in Fig. 1(b), during each transmission event, the system emits n consecutive plane waves with varying tilt angles and Hadamard-encoded orthogonal polarities at a fixed time delay interval. In this study, the time interval between adjacent pulses is set to 1 μs. A pulse interval that is too long reduces the imaging frame rate, while one that is too short causes signal interference due to the tailing effect.

During the reception phase, the transducer captures the superimposed echoes of n plane waves. Thanks to the orthogonality of Hadamard coding, the received echo data can be decoded through matrix inversion. Specifically, all received echo signals are linearly weighted and combined using the transpose of the coding matrix, enabling the separation of echo data for each angle and achieving lossless decoding. Let the echo signal received for each transmission be denoted as $M_i$, and the set of n received echo signals as a combined vector $M = [M_1, M_2, \cdots, M_n]$. By multiplying this with the transpose of the coding matrix $H_n^t$, the RF echo signal for each angle $RF = [RF_1, RF_2, \cdots, RF_n]$ can be decoded, as described in[27]:

$$RF = H_n^t M. \quad (2)$$

Since the product of the coding matrix and its transpose forms an identity matrix of order n, the signal strength of each decoded independent scattering echo is n times greater than that obtained without coding. Subsequently, the first $P_i$ rows of the decoded signal $RF_i$ are discarded to ensure temporal alignment and eliminate the effects of the delay $\tau_i$:

$$P_i = \tau_i \cdot f_s, \quad (3)$$

where $f_s$ represents the echo sampling frequency.

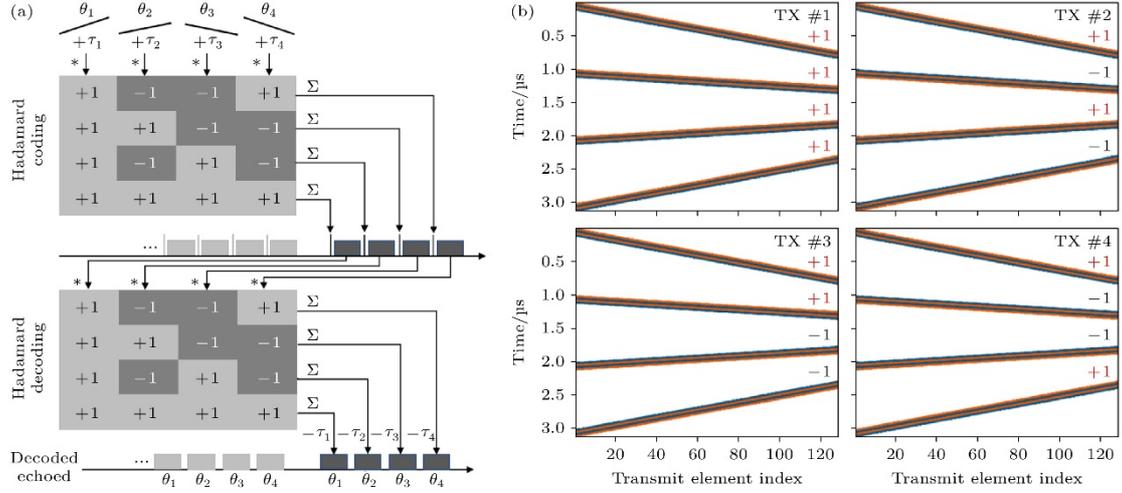

**Figure 1.** Hadamard-based multiplane wave encoding and decoding: (a) Schematic representation of the encoding and decoding process using a Hadamard matrix for four plane waves; (b) waveform examples for the four transmission events.

## 2.2 WLS based Vector Doppler velocity measurement

After the RF echo signal $RF_{\theta i}, i=1, 2,\cdots n$, of each angle is obtained by decoding, the echo signal of each angle is independently subjected to delay and sum (DAS) beamforming, in-phase/quadrature (I/Q) demodulation and tissue clutter filtering based on singular value decomposition (SVD) to extract the dynamic blood flow signal $S_{\theta i}$. Then the velocity projection $V_{\theta i}$ [28] of each pixel point at each angle is obtained based on the delay autocorrelation method, that is,

$$R_{\theta i} = \frac{1}{Nt-1}\sum_{t=1}^{Nt-1} S_{\theta i}(t)\cdot S_{\theta i}(t+1), \tag{4}$$

$$V_{\theta i} = \frac{-c\cdot PRF}{4\pi n f_0}\arg(R_{\theta i}), \tag{5}$$

Where $Nt$ represent the number of frames required to reconstruct a flow Doppler image, $R_{\theta i}$ represents the autocorrelation matrix, $c$ represents the speed of sound in the tissue, PRF represents the pulse transmission frequency, and $f_0$ represents the center frequency of the transmission signal. The geometric relationship between the velocity vector and the velocity projection $V_{\theta i}$ at each angle [23], can then be constructed

$$V_x\cdot\sin(\theta_i)+V_z\cdot(\cos(\theta_i)+1)=2\cdot V_{\theta i}, \tag{6}$$

Where $V_x$ and $V_z$ represent the horizontal and vertical projection components of the velocity vector in the plane rectangular coordinate system, respectively. Equation (6) is expressed in matrix form:

$$\begin{bmatrix} \sin\theta_1 & \cos\theta_1+1 \\ \sin\theta_2 & \cos\theta_2+1 \\ \vdots & \vdots \\ \sin\theta_n & \cos\theta_n+1 \end{bmatrix} \times \begin{bmatrix} V_x \\ V_z \end{bmatrix} = 2 \times \begin{bmatrix} V_{\theta 1} \\ V_{\theta 2} \\ \vdots \\ V_{\theta n} \end{bmatrix}. \tag{7}$$

The equation is overdetermined and have no exact solution. The conventional method is to use the least square method to solve the equations. However, considering that the measurement data at different angles may be affected by different degrees of noise, we use the WLS method to solve the velocity vector:

$$A = \begin{bmatrix} \sin\theta_1 & \cos\theta_1+1 \\ \sin\theta_2 & \cos\theta_2+1 \\ \vdots & \vdots \\ \sin\theta_n & \cos\theta_n+1 \end{bmatrix}, \tag{8}$$

$$\begin{bmatrix} V_x \\ V_z \end{bmatrix} = 2 \cdot \left(A^T W A\right)^{-1} A^T W \begin{bmatrix} V_{\theta 1} \\ V_{\theta 2} \\ \vdots \\ V_{\theta n} \end{bmatrix}, \tag{9}$$

where $W$ is a diagonal weight matrix, and each diagonal element $w_{\theta i}$ corresponds to the weight of the residual at a certain angle. The size of the weight value is optimized by iteration, keeping the sum of the weight values at 1, and initially setting all the weight values of $w_{\theta i}$ to $1/n$, that is,

$$W = diag(w_{\theta 1}, w_{\theta 1}, \cdots, w_{\theta n}). \tag{10}$$

Calculate the residual $res_{\theta i}$ at each angle as

$$res_{\theta i} = 2V_{\theta i} - V_z \cdot (\cos(\theta_i)+1) - V_x \cdot \sin(\theta_i). \tag{11}$$

The weight values are then updated according to the residual size:

$$w_{\theta i} = \frac{1}{|res_{\theta i}| + \lambda}, \tag{12}$$

Where $\lambda$ is the regularization coefficient to avoid the denominator being 0, and $\lambda$ is set to 0.01 here. After the updated weight is normalized, the next iteration is continued until convergence, that is, the iteration is stopped when the change of the weight is less than the preset threshold $\varepsilon$:

$$\max \left| w_{\theta i}^{(k+1)} - w_{\theta i}^{(k)} \right| < \varepsilon, \tag{13}$$

Where $k$ is the number of iterations. Through iteration, we finally get the estimation results of the horizontal velocity component $V_x$ and the vertical velocity component $V_z$ of each pixel. The weight matrix $W$ optimized by iteration reflects the contribution of the measurement data to the estimation of the velocity vector at each angle. The total velocity magnitude $V_{UVD}$ and direction $\theta_{UVD}$ at each pixel location can then be calculated:

$$V_{UVD} = \sqrt{(V_x^2 + V_z^2)} \:, \tag{14}$$

$$\theta_{UVD} = \tan^{-1}(\frac{Vx}{Vz}) \:. \tag{15}$$

By analyzing the pulsatility characteristics of vector velocity magnitude $V_{UVD}$ during the cardiac cycle, the peak systolic velocity (PSV) and end-diastolic velocity (EDV) can be extracted. The RI [29] for each pixel location can then be calculated:

$$RI = \frac{PSV - EDV}{PSV} \:. \tag{16}$$

**2.3 Spiral Flow Simulation Setup**

The probe parameters in the simulation match those used in the rat experiment, with the probe model being L22-14vX (Vermon, center frequency 18 MHz, 128 elements). The pulse duration is 2 cycles, the sampling frequency is 65 MHz, and the sound speed is 1540 m/s. An 8th-order Hadamard matrix is utilized for encoding and decoding, with a pulse repetition frequency of 14.4 kHz, 8 tilt angles (-15° to 15°), resulting in a composite frame rate of 1.8 kHz; 100 composite frames were collected. Band-limited white noise (-10 dB, 1–22 MHz) is introduced to the analog channel data before decoding to simulate the specified effective blood noise contrast. To minimize computational load, the vessel wall and surrounding tissue are excluded from the simulation, and thus no clutter filtering is applied.

To replicate the laminar flow characteristics of spiral blood flow velocity, the geometry of the spiral blood vessel is first established. The vessel midline follows an Archimedes spiral, defined by the following polar coordinate equation:

$$r(\alpha) = a + b\alpha \:, \tag{17}$$

where $a$ represents the initial radius of the spiral, $b$ governs the spacing between spiral turns, and $\alpha$ ranges from 0 to $\alpha_{max}$. We set a = 1 mm, b = 0.28, and $\alpha_{max}$ = 4.2π. The equations for the vessel walls on either side are

$$r_1(\alpha) = a + b\alpha - R \:, \tag{18}$$

$$r_2(\alpha) = a + b\alpha + R, \tag{19}$$

where $R$ is the vessel radius, fixed at 0.6 mm. To simulate scatterers in the blood flow, 300 scatter points are assigned per square millimeter. The scatter points have random intensity (mean of 0, standard deviation of 1), and the velocity of each scatter point relates to its distance from the vessel midline as follows:

$$u = u_c[1 - (l/R)^2], \tag{20}$$

where $u_c$ denotes the maximum blood flow velocity, set at 25 mm/s, and $l$ is the distance from the scatter point to the vessel center, ranging from -R to R. This flow pattern is laminar, aligning with the characteristics of low-speed blood flow in small vessels. The velocity of each scatter point dictates the change in its angle α per frame (see Appendix for derivation):

$$\frac{d\alpha}{dt} = \frac{u}{\sqrt{b^2 + (a + b\alpha + l)^2}}, \tag{21}$$

With an imaging frame rate of 1000 Hz, the time step $\Delta t$ is 1/1000 s. To keep scatter points within the defined spiral range, any point exceeding the maximum angle $\alpha_{max} = 4.2\pi$ is reset to the starting position. Finally, polar coordinates $(r, \alpha)$ are converted to rectangular coordinates $(x_p, z_p)$:

$$x_p = r\cos\alpha + x_{offset}, z_p = r\sin\alpha + z_{offset}, \tag{22}$$

where $x_{offset} = 0$ mm and $z_{offset} = 7$ mm denote the horizontal and vertical coordinate offsets, respectively.

The simulation is conducted using Field II (Release 3.30) software, which simulates and generates realistic ultrasonic signals and establishes a propagation model[30,31]. All simulations and data analyses were performed on a workstation equipped with an Intel Core i7-12700K processor and 32 GB of memory, using MATLAB (R2022a, MathWorks, USA) software. Simulating a single-angle plane wave echo takes approximately 93 seconds, with a total computation time of about 21 hours for a full simulation of 100 composite frames (each derived from eight oblique-angle plane wave echoes).

## 2.4 Experimental setting of cerebral blood flow imaging in rats

The animal experiment was approved by the Animal Welfare and Ethics Committee of the Department of Animal Experimental Science at Fudan University (Approval No.: 202202020Z). Male Sprague-Dawley rats, aged 8 weeks and weighing approximately 200 g, were used in the experiment. The rats were continuously anesthetized with 2.5% isoflurane gas during the experiment. Prior to ultrasound imaging, a rectangular cranial window (bregma +4 mm to -8 mm, width 13 mm) was created at the top of the rat skull. The rat skull roof was shaved, the skin was disinfected with iodophor, and an incision was made in the skull roof using a scalpel. The skin and periosteum were removed using surgical scissors to expose the skull. After thinning the skull with a dental drill, it was carefully removed, leaving the dura mater intact. During imaging, the rat's head was secured with a stereotaxic instrument, body temperature was maintained at 39°C using a constant-temperature heating pad, the ultrasound probe was positioned directly above the coronal plane of the brain (bregma -5.6 mm), and the probe was coupled to the rat brain with a medical ultrasound coupling agent. Multi-channel ultrasonic data transmission and acquisition were performed using a custom-developed programmable phased ultrasound experimental platform. The probe used in the experiment was identical to that described in the simulation setup. An 8th-order Hadamard matrix was employed to encode and decode the pulses, with a pulse repetition frequency of 14.4 kHz, 8 tilt angles (-15° to 15°), a composite frame rate of 1.8 kHz, an acquisition duration of 0.2 s, and a total of 360 composite frames; see Table 1 for detailed parameter settings.

Table 1. Overview of parameters for simulation and rat experiments

| Parameters | Simulation Exp. | Rat Exp. |
|---|---|---|
| Center Frequency/MHz | 15.625 | 15.625 |
| PRF/kHz | 14.4 | 14.4 |
| Number of Angles | 8 | 8 |
| Compounded Frame Rate/kHz | 1.8 | 1.8 |
| Acquisition Duration/s | 1/18 | 0.2 |
| Total Acquired Frames | 100 | 360 |
| Number of Encoded Pulses | 8 | 8 |
| Pulse Interval/μs | 1 | 1 |
| Voltage/V | \ | 20 |

## 3. Results of Simulation Experiments

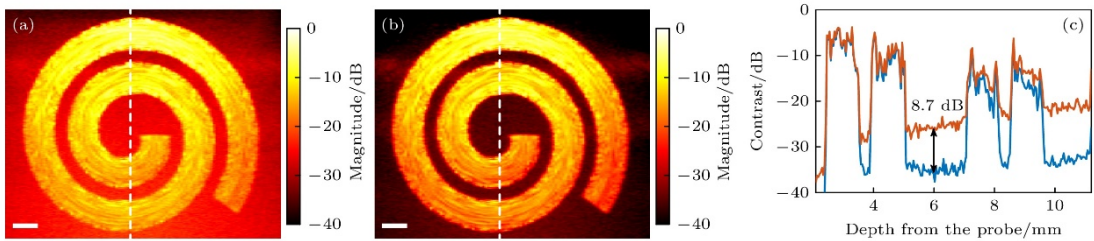

**Figure 2.** Simulated results of spiral flow power Doppler imaging based on conventional ultrafast ultrasound power Doppler (PD) and pulse coded power Doppler (PC-PD): (a) PD simulation result; (b) PC-PD simulation result; (c) SNR quantization curve, showing the contrast of blood flow and background noise at different depths.

Figure 2 presents the simulation results of spiral blood flow imaging using ultrafast power Doppler (PD) and pulse-coded power Doppler (PC-PD). In Fig. 2(a), the blood flow signal exhibits noticeable attenuation, with progressively reduced blood flow-to-noise contrast, making it challenging to discern vessel edges at greater imaging depths. A comparison of Fig. 2(a) and Fig. 2(b) reveals that PC-PD significantly enhances imaging quality and SNR compared to conventional PD mode. Fig. 2(c) displays the quantitative SNR analysis for both imaging modes at various depths, measured along the white dotted lines in Fig. 2(a) and Fig. 2(b). The vertical axis indicates signal contrast, the horizontal axis denotes imaging depth, with the orange curve representing conventional PD mode and the blue curve representing PC-PD mode. The comparison demonstrates that the SNR of PC-PD mode exceeds that of PD mode by approximately 8.7 dB across the entire imaging depth range (0–11 mm), suggesting that Hadamard matrix-based pulse coding and decoding effectively reduces noise intensity.

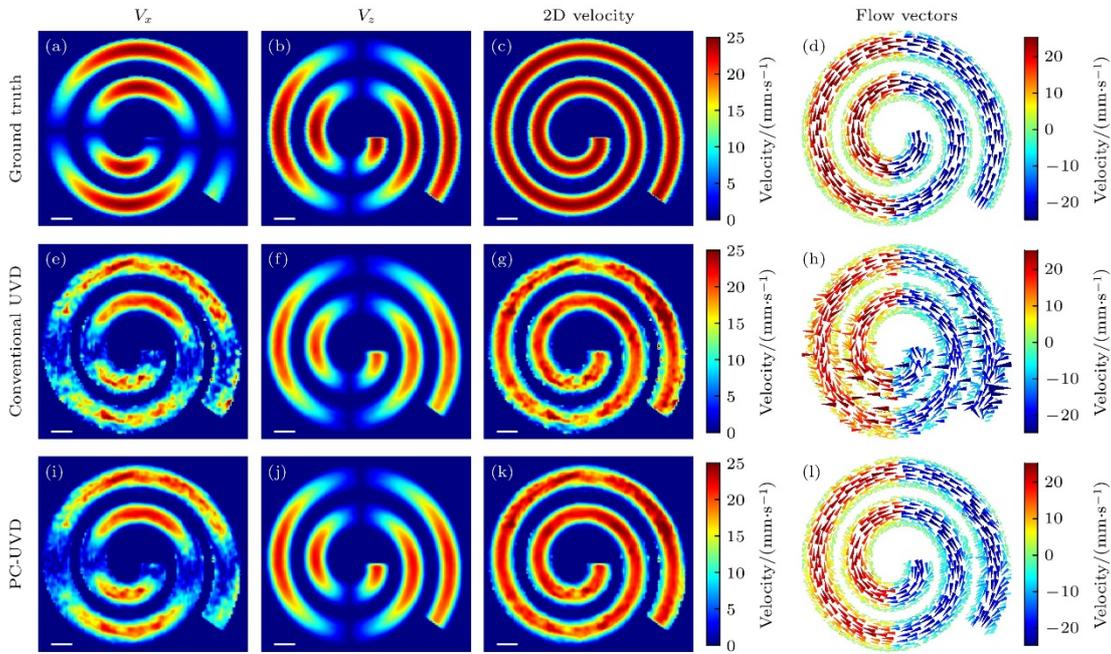

**Figure 3.** Simulated results of vector velocity in spiral blood flow, including the ground velocity truth (top row), conventional UVD (middle row), and PC-UVD (bottom row) for blood flow velocity. Each row contains the horizontal velocity component $V_x$ (first column), vertical velocity component $V_z$ (second column), 2D velocity (third column), and vectorized blood flow imaging results (fourth column), respectively.

Figure 3 presents the theoretical true velocity of spiral blood flow alongside vector velocity simulation results from conventional UVD and the proposed PC-UVD method. Fig. 3(a)–(d) depict the theoretical true velocity of the spiral blood flow. As shown in Fig. 3(c), the theoretical model's two-dimensional velocity map accurately depicts blood flow velocity intensity distribution, while the vector map in Fig. 3(d) precisely illustrates velocity magnitude and direction, clearly highlighting spiral flow field characteristics. Fig. 3(e)–(h) present velocity estimation results from the conventional UVD method. Compared to the theoretical true value, conventional

UVD measurements exhibit greater error. In Fig. 3(e), the horizontal velocity component Vx displays significant noise interference and measurement errors, particularly in deeper regions with low SNR. Since Doppler measurements are more sensitive to axial motion, the vertical velocity component Vz (Fig. 3(f)) shows better results than Vx, with a smoother velocity distribution and no prominent outliers, though peak estimated velocity is slightly below the theoretical true value. The accuracy of the two-dimensional velocity map (Fig. 3(g)) is substantially compromised, and while the velocity vector map (Fig. 3(h)) generally reflects the spiral flow pattern, notable anomalous measurements are evident in deeper areas with low SNR. Fig. 3(i)–(l) display simulation results from the proposed PC-UVD method. As shown in Fig. 3(i), compared to conventional UVD results, the proposed method provides more accurate horizontal velocity estimates, with significantly reduced measurement error in deeper areas and a smoother velocity distribution. The vertical velocity component Vz (Fig. 3(j)) and the two-dimensional velocity map (Fig. 3(k)) more closely align with the theoretical true value. The velocity vector map in Fig. 3(l) indicates that PC-UVD results exhibit no significant vector velocity anomalies, with overall velocity distribution and direction closely matching the theoretical model.

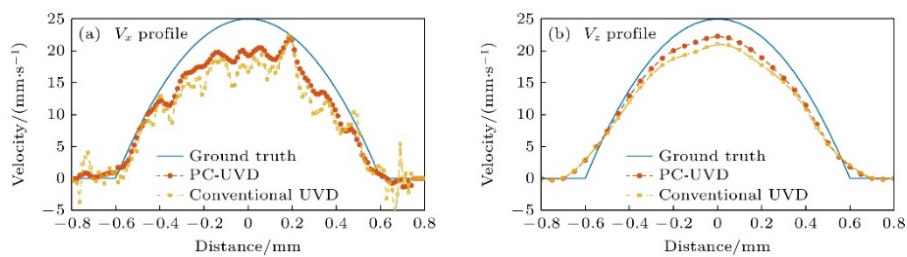

**Figure 4.** Horizontal and vertical velocity profiles of spiral blood flow measured by ultrafast vector Doppler: (a) Lateral velocity distribution; (b) axial velocity distribution. The blue line represents the velocity ground truth, the yellow line represents the conventional UVD measurement, and the red line represents the PC-UVD measurement.

Fig. 4 shows the distribution of the measured horizontal and vertical blood flow velocities versus the distance from the center of the vessel. In order to simulate the laminar flow mode, in the simulation setting, we set the velocity magnitude to be distributed according to the parabolic law, as shown by the blue line in Fig. 4. The maximum velocity in the midline of the vessel is 25 mm/s, and the velocity near the vessel wall is 0. It can be seen from Fig. 4(a) that although the transverse blood flow velocity measured by conventional UVD and PC-UVD is slightly lower than the theoretical velocity in the whole range, the transverse blood flow velocity measured by the latter is closer to the theoretical value, and the random fluctuation affected by noise is well suppressed. Fig. 4(b) reflects that the measurement accuracy of the vertical component of the blood flow velocity by the PC-UVD method is also relatively improved.

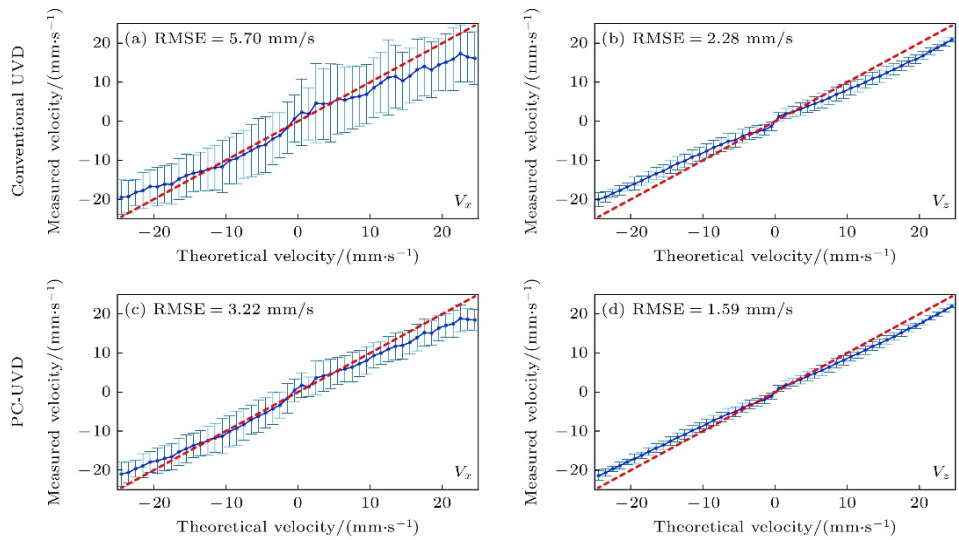

**Figure 5.** Accuracy analysis of flow vectors derived from the spiral flow phantom. Plots of estimated velocity versus ground-truth velocity in (a), (c) lateral and (b), (d) axial directions, as extracted from the velocity values at different pixel positions within the spiral loop; error bars denote standard deviation over a minimum of 100 measurements. The RMSE of all pixel position measurements is demonstrated in the upper left corner of each of measurements at all pixel positions is displayed in the upper left corner of each plot: (a), (b) Analysis of results of conventional UVD; (c), (d) analysis of results of PC-UVD.

The accuracy of global vector velocity estimation is quantitatively analyzed. Fig. 5(a), (b) and Fig. 5(c), (d) are the analysis charts of velocity estimation error for conventional UVD and PC-UVD, respectively. Fig.5(a), (c) and Fig. 5(b), (d) respectively plot the estimated horizontal and vertical components of the velocity against the standard value. It is worth noting that the greater the velocity, the greater the deviation of the estimate. All measurement error values within 1 mm/s are used to calculate a standard deviation and are shown with error bars. Each error bar represents the standard deviation of a minimum of 100 measurements. The root-mean-square error (RMSE) of all pixel position measurements is shown in the upper left corner of the figure. For horizontal velocity estimation, the measured RMSE of conventional UVD is 5.7 mm/s, the measurement RMSE of PC-UVD is reduced to 3.22 mm/s, and the accuracy is improved by 43.51%. For vertical velocity estimation, the measurement RMSE of conventional UVD is 2.28 mm/s, the measurement RMSE of PC-UVD is reduced to 1.59 mm/s, and the accuracy is improved by 30.26%.

## 4. Results of In-vivo Experiments

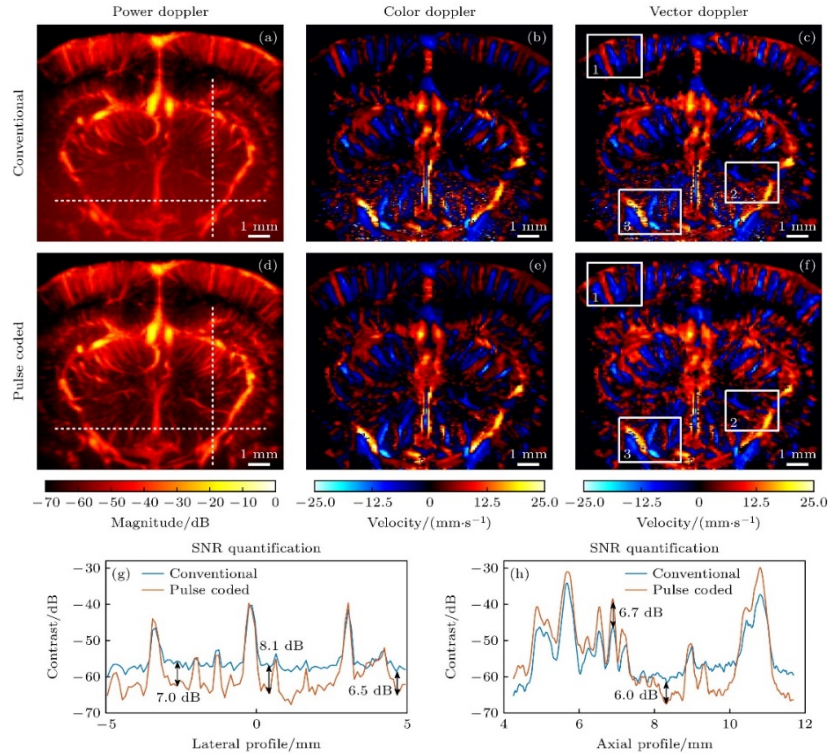

**Figure 6.** Ultrafast Doppler images of rat cerebral blood flow. (a)–(c) shows the ultrafast (a) power Doppler, (b) color Doppler, and (c) vector Doppler results obtained in conventional plane wave transmission mode. (d)–(f) presents the ultrafast (d) power Doppler, (e) color Doppler, and (f) vector Doppler results obtained in pulse coded plane wave transmission mode. In the ultrafast color Doppler and vector Doppler images, red and blue indicate upward and downward blood flow directions, respectively. (g) Signal intensity profile along the horizontal dashed line in (a) and (d); (h) signal intensity profile along the vertical dashed line in (a) and (d).

Fig. 6 is the ultrafast Doppler coronal section (bregma –5.6 mm) of rat cerebral blood flow. Fig. 6(a) and Fig. 6(d) are the ultrafast power Doppler blood flow images of plane wave conventional emission mode and pulse coding emission mode, respectively. Compared with Fig. 6(a), the contrast between small vessels and background noise in Fig. 6(d), especially in the deep region, is significantly improved, and the number of small vessels that can be detected is also increased. In order to quantitatively analyze the SNR enhancement caused by pulse coding, Fig. 6(g) and Fig. 6(h) show the signal intensity profiles at the horizontal and vertical dotted lines in Fig. 6(a) and Fig. 6(d), respectively. The results show that the SNR of ultrafast Doppler flow imaging based on pulse coding mode is improved by about 6-8 dB. Fig. 6(b) and Fig. 6(e) are color Doppler flow images of plane wave conventional emission mode and pulse coding emission mode, respectively. Compared with Fig. 6(b), the red and blue random noise in the deep brain region of Fig. 6(e) is significantly suppressed. Fig. 6(c) and Fig. 6(f) are conventional UVD and PC-UVD velocity maps, respectively. Similar to color Doppler, vector velocity values are mapped to red and blue according to the upper and lower directions of the axial velocity component.

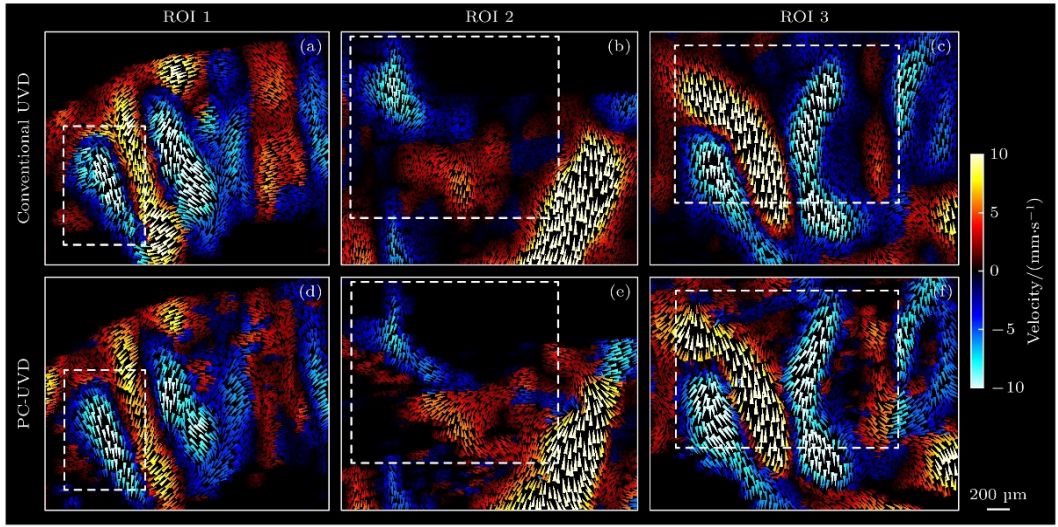

**Figure 7.** Ultrafast vectorized Doppler velocity imaging results in three ROIs in the rat brain. The three ROIs correspond to the three rectangular areas in Fig. 6(c) and Fig. 6(f). (a)–(c) shows the velocity measurements based on conventional UVD, while (d)–(f) shows the results based on PC-UVD. The velocity vectors are represented by small triangles, with the area and color indicating the magnitude of the blood flow velocity, and the sharp angles pointing in the direction of the flow.

Compared with color Doppler, which can only measure the axial velocity component of blood flow, vector Doppler can measure the blood flow velocity more accurately. Three brain regions of interest (white boxes) were selected at different depths in the blood flow velocity vector map (Fig. 6(c) and Fig. 6(f)) for velocity vector visualization (Fig. 7). The velocity vector was represented by a small triangle, the area and color of the triangle represented the blood flow velocity, and the direction pointed by the acute angle of the triangle was the direction of the blood flow velocity. The conventional UVD method (Fig. 7(a-c)) still has some measurement limitations in the complex in vivo small vascular environment. For example, the velocity vector in the dotted line box in Fig. 7(a) has an obvious error of "hitting the vessel wall", which reflects that the measurement of the magnitude and direction of the blood flow velocity near the vessel wall is not accurate; the imaging SNR of the deep brain in Fig. 7(b) is insufficient, the small vessels are obviously blurred, and some vessels are missing; the measurement of the transverse component of the blood flow vector in Fig. 7(c) is obviously wrong, which makes the estimated blood flow direction inconsistent with the direction of the vessels. The above limitations were significantly improved in PC-UVD measurements (Fig. 7(d-f)).

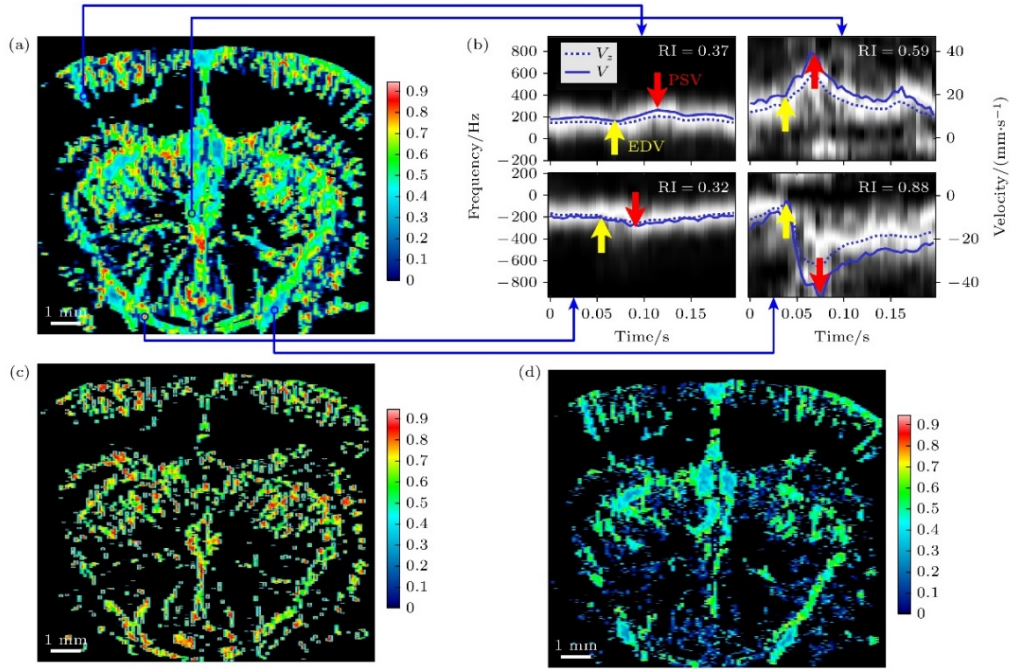

**Figure 8.** PC-UVD based rat cerebral blood flow resistivity index measurements: (a) Cerebral blood flow resistivity index map; (b) multi-angle compounded Doppler spectra at different locations throughout a cardiac cycle. The dashed lines overlaid on the spectrum indicate the center frequency of the Doppler signal, while the solid lines represent the measured vector Doppler dynamic velocities. The red arrows point to the peak systolic blood flow velocity, and the yellow arrows point to the end-diastolic velocity. Top left: venous flow upward; bottom left: venous flow downward; top right: arterial flow upward; bottom right: arterial flow downward. (c) Arterial flow map. (d) Venous flow map.

By analyzing the changes of cerebral blood flow vector velocity measured by PC-UVD in a cardiac cycle, the blood flow pulsation characteristics of all pixels in the whole brain were extracted, and the blood flow RI value was further calculated. As shown in Fig. 8(a), the flow resistivity map based on ultrafast vector Doppler technique shows the RI distribution of different vessels. Fig. 8(b) shows the flow Doppler spectrum after multi-angle coherent recombination at different positions, and extracts the maximum systolic velocity and minimum end-diastolic velocity in each cardiac cycle. The dotted line in the spectrum diagram represents the center frequency of the Doppler spectrum after coherent recombination, which corresponds to the dynamic velocity estimation of ultrafast color Doppler, and the solid line represents the measured vector Doppler dynamic velocity. Both of them have synchronous fluctuation characteristics, which proves the accuracy of the dynamic estimation of ultrafast vector Doppler velocity. The red arrow in Fig. 8(b) gives the peak systolic velocity of the flow, while the yellow arrow gives the minimum end-diastolic velocity. The upper left and lower left panels show the dynamic velocity characteristics of up-flowing and down-flowing venules, respectively, and the upper right and lower right panels show the dynamic velocity characteristics of up-flowing and down-flowing arterioles. Fig. 8(c) and Fig. 8(d) show the arteriolar and venular flow networks, respectively. The resistivity of arterioles is generally between 0.55 and 0.9, while the resistivity of venules is generally less than 0.5[29].

## 5. Discussion

In this paper, an ultrafast pulse-coded vector Doppler blood flow velocity and resistivity imaging method is proposed, which can significantly improve the SNR of RF echo data and reduce the outliers of velocity estimation at all angles by virtually increasing the amplitude of the transmitted signal without affecting the imaging frame rate and imaging resolution, and adaptively reduce the weight of outliers by using weighted least squares estimation and residual iterative optimization, thus enhancing the adaptability to noise. In this paper, the effectiveness of PC-UVD method for dynamic estimation of small blood flow velocity is verified by spiral blood flow simulation experiment and in vivo imaging experiment of rat cerebral small vessel network, especially in the deep area with serious signal attenuation and low SNR, which shows obvious advantages over the conventional UVD method. The following discussion focuses on the advantages, limitations, and potential applications of the PC-UVD method.

In order to meet the requirement of high SNR for vector Doppler measurement, a transmission sequence based on cascade transmission of $n$ tilted plane waves is implemented with Hadamard amplitude coding. Each received echo signal contains the scattered echo of each inclined plane wave. Based on the orthogonality of Hadamard matrix, the echo signals of each angle can be decoded by simply weighting and compounding the echo signals received for $n$ times. In addition, the decoded signal strength is enhanced by $n$ times compared with the echo signal acquired by a single transmission, which is equivalent to virtually amplifying the probe transmission voltage to $\sqrt{n}$ times, and the SNR can be theoretically increased by $20 \cdot \log(\sqrt{n})$ dB[32]. $n=8$ is used in both the simulation and the in vivo experiment, so the theoretical SNR can be enhanced by 9.03 dB, and the SNR gain measured in the simulation experiment is about 8.7 dB, which is close to the theoretical value. In the rat in vivo experiment, the measured SNR gain was 6~8 dB, which was also consistent with the theoretical value. The order $n$ of a Hadamard matrix can be 2 or a multiple of 4, i.e. $n$ = 2, 4, 8, 12, 16, 20, 24, · · ·, etc. Although in theory, the larger $n$ is, the more significant the SNR gain is, the imaging frame rate and time resolution limitations determine that the value of $n$ should not be too large. In terms of imaging frame rate, because $n$ is not only equal to the number of pulses in a transmission event, but also equal to the number of angles of the inclined plane wave, with the increase of $n$, the number of angles increases accordingly, and the maximum composite imaging frame rate will decrease inversely. However, small flow imaging requires a high frame rate to meet high sensitivity and avoid spectral aliasing, so the value of $n$ should not be too large. In terms of imaging resolution, the premise of Hadamard lossless decoding is that the measured object remains quasi-static during $n$ emission and acquisition events. However, the blood flow velocity is dynamic, especially when measuring the blood flow resistivity, too large value of *n* may cause large velocity estimation error.

Imaging methods based on multi-pulse coded excitation also face safety challenges in clinical practice. The main safety indicators of ultrasound testing include mechanical index, thermal index, time-average peak intensity and pulse-average peak intensity. Although directly increasing the transmit voltage can improve the SNR, it will increase all the safety indicators at the same time. In contrast, multi-pulse coded transmission can virtually increase the transmit voltage without changing the mechanical index and the average peak intensity of the pulse, thus improving the safety. However, due to the increase of the number of pulses transmitted per unit time, the heat index and the time-averaged peak intensity will still increase. Therefore, the frame rate of signal transmission should be set reasonably to meet the safety requirements in clinical application.

Compared with the traditional focused scanning mode, ultrafast ultrasound significantly improves the imaging frame rate, but in high-speed blood flow imaging, the problem of frequency aliasing still needs to be overcome to ensure the accuracy of blood flow velocity estimation. In recent years, researchers have proposed a variety of solutions to overcome the problem of frequency aliasing. In 2016, Posada et al.[33] proposed an anti-aliasing method based on staggered pulse repetition frequency, which uses two different imaging frequencies to obtain two Doppler velocity values with different aliasing multiples, and then solves the Doppler velocity without aliasing. Approximately, in 2021, Poree et al.[34] proposed a dual-wavelength anti-aliasing method to solve Doppler velocities with different aliasing multiples through different wavelength combinations. In 2022, Ecarlat et al.[35] further improved the dual-wavelength method by using a broadband chirped wave instead of a sinusoidal pulse wave. In addition, deep learning methods have been introduced in recent years to effectively overcome the spectral aliasing problem in color Doppler[36] and vector Doppler[37] imaging.

Different with color Doppler, which can only measure a single component of blood flow velocity (the direction of sound wave propagation), the two-dimensional vector Doppler method studied in this paper can obtain the in-plane component of blood flow velocity, which is suitable for the case where the blood vessel is parallel to the linear array probe, such as the coronal section of the brain in this paper. However, for complex small vascular networks, such as vascular velocity measurement around tumors, further development of three-dimensional ultrafast vector Doppler imaging methods is needed in the future. At present, researchers have realized the three-dimensional ultrafast power and color Doppler imaging method based on two-dimensional phased array[32] and row-column addressing[38] array. Therefore, it is feasible and valuable to continue to develop the three-dimensional ultrafast vector Doppler imaging (3D-UVD) method on this basis.

## 6. Conclusion

In this paper, an ultrafast ultrasound vector Doppler blood flow velocity and resistivity measurement method based on Hadamard multi-pulse coding transmission is proposed, which can significantly improve the SNR of blood flow imaging and the accuracy of blood flow velocity measurement through spiral blood flow simulation experiments. Compared with the conventional UVD method, the SNR of the proposed method is improved by 8.7 dB, the velocity measurement error is reduced by more than 30%, and the abnormal estimation of velocity caused by insufficient SNR is effectively reduced. Cerebral blood flow in rats in vivo experiments show that the proposed PC-UVD method has the advantages of large imaging field of view, high SNR and high temporal and spatial resolution. It can effectively measure the fluctuation characteristics of global cerebral blood flow in a single cardiac cycle and generate flow resistivity index map, which has certain clinical value for the distinction of cerebral arterioles and veins and the early diagnosis and evaluation of cerebrovascular diseases.

## Appendix

In the polar coordinate system, the formula for the arc length of a curve is

$$ds = \sqrt{(\frac{dr}{d\alpha})^2 + r^2}\, d\alpha, \tag{A1}$$

Where $r(\alpha)$ is a function of radius as a function of angle. For any particle's helical trajectory:

$$r(\alpha) = a + b\alpha + l, \tag{A2}$$

Where $a$ is the radius of the starting point of the spiral; $b$ controls the spacing between the spirals; and $l$ represents the distance between the scattering point and the center of the vessel. Combining (A1) and (A2), we get:

$$ds = \sqrt{b^2 + (a + b\alpha + l)^2}\, d\alpha. \tag{A3}$$

The differential relation of arc length $s$ and velocity $u$ is

$$ds = u \cdot dt. \tag{A4}$$

Substituting (A3) into (A4), we can get:

$$\frac{d\alpha}{dt} = \frac{u}{\sqrt{b^2 + (a + b\alpha + l)^2}}. \tag{A5}$$